\documentclass[8.5pt,twoside,twocolumn]{article}
\usepackage{amsmath,amssymb}
\usepackage[super,sort&compress,comma]{natbib}
\usepackage[version=3]{mhchem}
\usepackage{times,mathptmx}
\usepackage{sectsty}
\usepackage{balance}

\usepackage{graphicx} 
\usepackage{lastpage}
\usepackage[format=plain,justification=raggedright,singlelinecheck=false,font=small,labelfont=bf,labelsep=space]{caption}
\usepackage{fancyhdr,xcolor,ulem}
\usepackage{subfigure} 
\usepackage[separate-uncertainty=true]{siunitx}

\usepackage{arydshln}

\begin{document}


\makeatletter
\def\subsubsection{\@startsection{subsubsection}{3}{10pt}{-1.25ex plus -1ex minus -.1ex}{0ex plus 0ex}{\normalsize\bf}}
\def\paragraph{\@startsection{paragraph}{4}{10pt}{-1.25ex plus -1ex minus -.1ex}{0ex plus 0ex}{\normalsize\textit}}
\renewcommand\@biblabel[1]{#1}
\renewcommand\@makefntext[1]%
{\noindent\makebox[0pt][r]{\@thefnmark\,}#1}
\makeatother
\renewcommand{\figurename}{\small{Fig.}~}
\sectionfont{\large}
\subsectionfont{\normalsize}


\twocolumn[
  \begin{@twocolumnfalse}
        \noindent\LARGE{\textbf{Surface patterns in drying films of silica colloidal dispersions}}
\vspace{0.6cm}

\noindent\large{\textbf{F.~Boulogne\textit{$^{a\dagger}$}$^{\ast}$,   F.~Giorgiutti-Dauphin\'e\textit{$^{a}$}, L.~Pauchard\textit{$^{a}$}}}\vspace{0.5cm}

\vspace{0.6cm}

\noindent \normalsize{
            We report an experimental study on the drying of silica colloidal dispersions.
            Here we focus on a surface instability occurring in a drying paste phase before crack formation which affects the final film quality.
            Observations at macroscopic and microscopic scales reveal the occurrence of the instability, and the morphology of the film surface.
            Furthermore, we show that the addition of adsorbing polymers on silica particles can be used to suppress the instability under particular conditions of molecular weight and concentration.
            We relate this suppression to the increase of the paste elastic modulus.
}
\vspace{0.5cm}
\end{@twocolumnfalse}
]

\footnotetext{
\textit{$^{a}$ UPMC Univ Paris 06, Univ Paris-Sud, CNRS, F-91405. Lab FAST, Bat 502, Campus Univ, Orsay, F-91405, France. Fax: +33 1 69 15 80 60; Tel: +33 1 69 15 80 46; E-mail: boulogne@fast.u-psud.fr\\
\textit{$^{\dagger}$ Now at: Department of Mechanical and Aerospace Engineering,
    Princeton University, Princeton, NJ 08544, USA.}
}}

\section{Introduction}

Patterns arising in soft materials such as elastomers, gels and biological tissues receive a growing attention \cite{Li2012}.
The understanding and the control of the underlying instabilities are crucial for technological applications (microelectronics, microfluidics), for biological systems (wrinkling of human skin or drying fruit \cite{Cerda2003}) or for medical applications where gels are used as biological scaffolds for tissues or organs.
In addition, various patterns which affects the surface of films are reported in the literature.
One of the most studied concerns wrinkles observed when a hard skin, sitting on a soft layer, is compressed.
Beyond a critical strain, it results in a periodic sinusoidal deformation of the interface \cite{Huraux2012,Xuan2012,Chen2012a,Genzer2006} for which the periodicity has been derived theoretically \cite{Cerda2003,Groenewold2001}.
When the strain is further increased, a secondary instability occurs leading to wrinkle-to-fold transition \cite{Brau2013,Brau2011}; this  results in a localization of the deformation \cite{Pocivavsek2008}.
Under a biaxial compressive stress, it has been shown that a repetitive wrinkle-to-fold transition produces a hierarchical network of folds \cite{Kim2011,Reis2011}.
These patterns are observed in various systems such as glassy polymers \cite{Ebata2012,Huraux2012}, polyethylene sheets \cite{Cerda2003} or foams \cite{Reis2009}.
Another group of patterns arises in soft layers supported by rigid substrates \cite{Li2012}.
Under compressive stresses, an instability often called creasing is manifested by localized and sharp structures at the free surface.
These are commonly observed in elastomers \cite{Hong2009,Cai2012}, and hydrogels \cite{Trujillo2008}, or in the situation of a rising dough in a bowl \cite{Cai2010}.
Despite similarities in their morphologies, origins of folds and creases are strongly different.
Whereas folds are a secondary instability of wrinkles, creases are a direct deformation of a flat film (creases are also known as sulci \cite{Hohlfeld2011,Tallinen2013}).
Moreover drying films can produce characteristic crack patterns as they are dried\cite{Xu2009,Routh2013}.
The drying-induced cracks can invade the surface and propagate simultaneously into the volume of the medium with evaporation of solvent, resulting in the division of the plane into polygonal domains.

In this paper, we experimentally investigate patterns displayed at the surface of drying films of silica nanoparticles in an aqueous solvent.
We discuss the occurrence of these patterns, and we examine their main features using different imaging techniques and rheological measurements.
Moreover, the polymer/silica interaction is usually used to tune the mechanical properties of composite films.
Consequently, we propose here a method to suppress these structures by adding adsorbing polymers to the silica nanoparticles.

\section{Experimental}\label{sec:exp_setup}
\begin{figure}
    \begin{center}
        \includegraphics[scale=0.9]{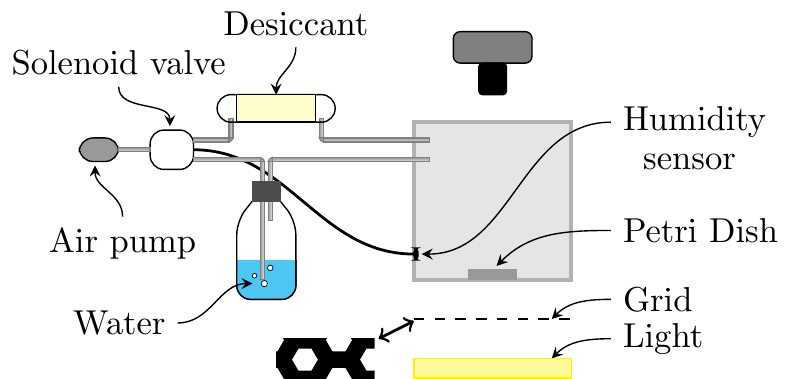}
    \end{center}
    \caption{Experimental setup. Dry or moist air is produced by an air flow from the ambient atmosphere through desiccant or water respectively to the box.
        Depending on the humidity measured by the humidity sensor inside the box, a solenoid valve is actioned to converge the humidity to the desired value.
        A honeycomb grid (represented in its actual size) is used to improve the contrast of the surface corrugations (Schlieren technique).
    }\label{fig:setup}
\end{figure}

\subsection{Controlled drying conditions}
Experiments consist in drying a colloidal suspension in a circular glass Petri dish (inner diameter $2R=5.6$ cm) placed in a chamber which has controlled temperature ($22\pm 2$ $^\circ$C) and relative humidity ($R_H=50\pm2$\%) \cite{Boulogne2013}.
The setup is sketched in figure \ref{fig:setup}.
Top and bottom walls of the chamber are transparent for visualization by transmitted light.
The bottom wall is carefully adjusted horizontally prior each experiment.
Since the sample contrast is very low, the Schlieren technique\cite{Weinstein2010} is used: a honeycomb grid is positioned between the bottom wall and an extended light source.
A camera (Nikon D300), located at the top of the chamber, records a photograph each minute.

In the following, $m_i$ varies in the range $1$ to $9$ g, resulting in a initial thickness $h_{i}$ varying in the range from $\simeq 0.3$ to $2.5$ mm.

\subsection{Colloidal dispersions}
\begin{table}
    \caption{Main properties of silica dispersions used in this paper.
    The volume fraction is noted $\phi_{0}$.
    Values of the particle diameters come from reference \cite{Bergna1994} (p.~324).}\label{tab:struc_silices}
    \centering
    \begin{tabular}{|c|c|c|c|c|}
        \hline
        Silica & diameter $2a$ (\si{\nano\meter}) & $\phi_{0}$ & $\rho$ (\si{\kilo\gram\per\meter\cubed}) & pH  \\
        \hline
        SM & 10 & 0.15 & 1180 & 9.9 \\
        HS & 16 & 0.19 & 1250 & 9.8  \\
        TM & 26 & 0.20 & 1260 & 9.2 \\
        \hline
    \end{tabular}
\end{table}

We use three aqueous dispersions of silica colloidal particles: Ludox SM-30, HS-40 and TM-50, commercially available from Sigma-Aldrich.
The pH is in the range of 9-10, and so the particle surface bears a high negative charge density \cite{Iler1979}.
SM-30 is used without treatments, while HS-40, and TM-50 are diluted using pure water (milliQ quality, resistivity: $18$ M$\Omega$.cm) at pH 9.5 by addition of NaOH.
A weight ratio of 90/10 (HS-40/water) and 75/25 (TM-50/water) are chosen to obtain similar initial volume fractions.
In the following, SM (unmodified), HS and TM designate these dispersions; their main properties are reported in Table \ref{tab:struc_silices}.

The effect of polymer chains on TM particles is investigated using polyethylene oxyde (PEO) or alternatively polyvinylpyrrolidone (PVP).
Indeed, a high affinity for silica surfaces is known for PEO and PVP resulting in an adsorption on silica particles.
To simplify the notations, polymers are noted $P_i$ as follow.
Different molecular weights are studied ($P_0$: \SI{300}{\dalton}, $P_1$: \SI{600}{\dalton}, $P_2$: \SI{3350}{\dalton}, $P_3$: \SI{6000}{\dalton}, $P_4$: \SI{35}{\kilo\dalton}, $P_5$: \SI{600}{\kilo\dalton} and $P_6$: \SI{8}{\mega\dalton}) for PEO and $P_7$: \SI{40}{\kilo\dalton} for PVP.
Polymers are purchased from Sigma-Aldrich and are dissolved in pure water at pH 9.5.
This solution, with a weight concentration of polymers noted $C_p$, and is used for the dilution of TM-50.
As a result, the final polymer concentration in TM dispersion is $C_p/4$.

The adsorption of polymers on silica particles has been largely studied in the literature \cite{Lafuma1991,Otsubo1990,Wong1992,Parida2006}.
In particular, it has been shown, from adsorption isotherms, that silica particles are totally covered with polymers for concentrations above $1$ \si{\milli\gram\per\meter\squared} for PEO \cite{Lafuma1991} and PVP \cite{Ilekti2000}.
In our experiments, the polymer concentrations do not exceed $\lesssim 0.01$ \si{\milli\gram\per\meter\squared}, which is much lower than the covering concentration.
In this concentration range, polymers adsorb on particles to form necklaces\cite{Wong1992}; the dispersed state is stable since the interactions between necklaces are repulsive\cite{Cabane1997}.

\subsection{Visualisation techniques}

Observation using an optical microscopy (DM2500 Leica microscope) by transmitted light is used at a macroscopic scale.

At a microscopic scale, the surface profile is investigated using an Atomic Force Microscope (AFM, Vicco) in tapping mode on dried TM samples (scanning area are $80\times80$ \si{\micro\meter\squared}) prepared with an initial weight $m_i=6$ g.
To complete these measurements at the onset of the propagation, an optical profiler is used (Taylor Hobson, $10\times$, working distance $7$ mm).
This contact-less method allows us to measure the surface profile of consolidating materials with a typical surface area $1.6\times1.6$ mm$^2$.

\section{Results}

\subsection{Macroscopic observations}

\subsubsection{Temporal evolution}

\begin{figure*}
    \includegraphics[width=\linewidth]{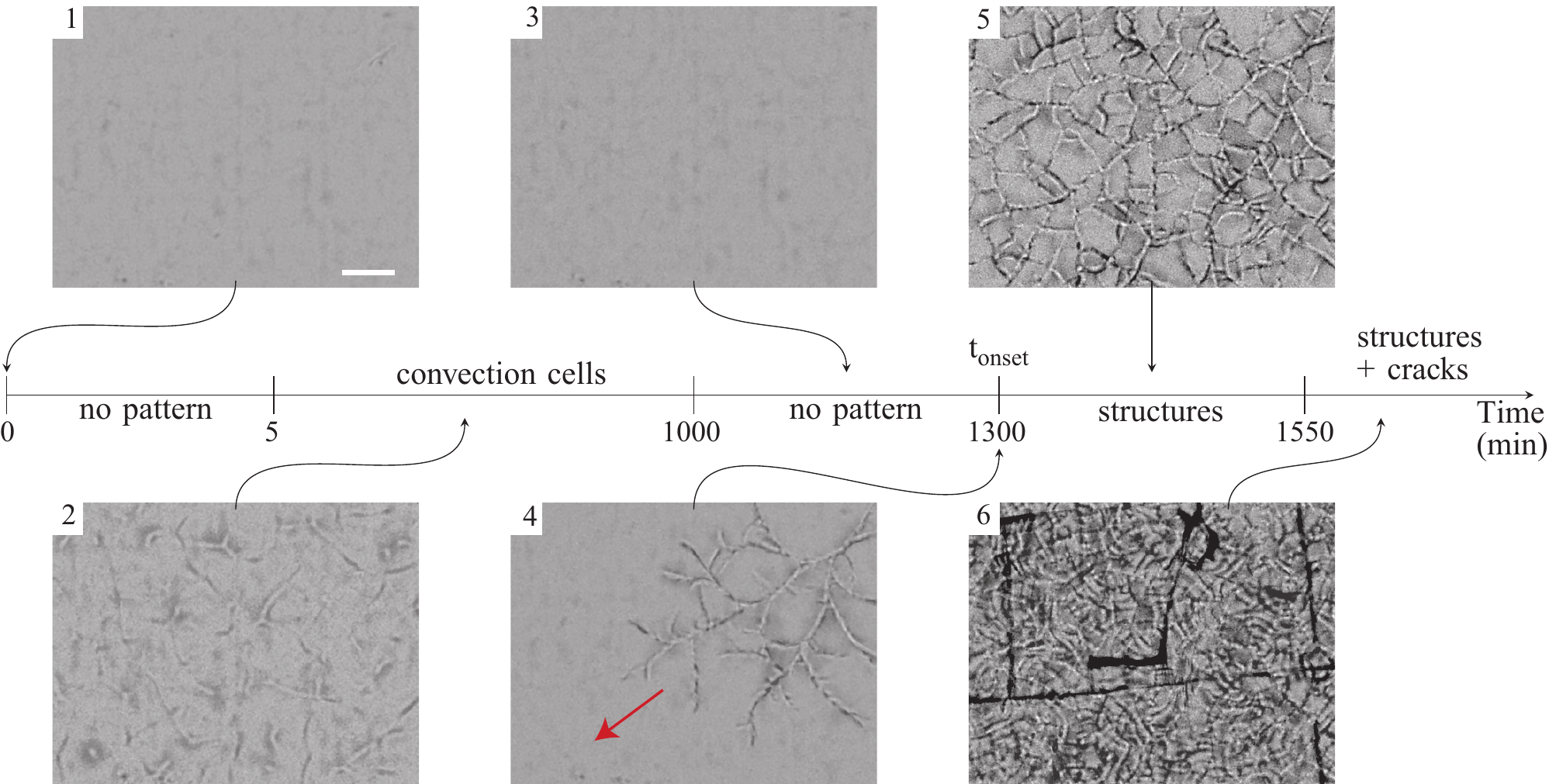}
    \caption{Time evolution of a flat film of a TM sample (initial weight is $m_i=6$ g).
        Each image is focused on the film surface (in the center, far from the petri dish edge) and is taken at the same region of the film.
        The arrow in image (4) gives the mean direction of propagation of the structures.
        Image (6) shows the final pattern made of sinuous structures superposed to a network of channeling cracks.
    Scale bar: $2$ mm.}\label{fig:timeline}
\end{figure*}

During water removal, particles concentrate and the film thickness decreases.
At a given time, a network of fine dark lines progressively invades the flat region of the film from the center to the edges.
The time evolution of the film surface is shown in figure \ref{fig:timeline}.
Starting from an homogeneous surface (image 1 in figure \ref{fig:timeline}), surface corrugation can be observed after a few minutes of evaporation (image 2 in figure \ref{fig:timeline}).
The corrugation patterns probably result from Rayleigh-B\'enard or B\'enard-Marangoni convective instabilities\cite{Bassou2009}.
However, the convective cells disappear after a period of time; the surface recovers its visual homogeneity (image 3 in figure \ref{fig:timeline}).
Then, at time $t_{onset}$, structures progressively invade the surface of the flat film (image 4 in figure \ref{fig:timeline}) and results in the pattern shown in image 5 in figure \ref{fig:timeline}.
Finally, the classical crack pattern forms in the film (image 6 in figure \ref{fig:timeline}).

\subsubsection{Onset time}

Starting from an homogeneous surface, the structures appear where the film is thinner, \textit{i.e.} in the center, far from the meniscus.
For TM dispersions, we observed that below a critical initial weight $m_i^c = 1.4\pm0.3$ g, \textit{i.e.}, $h_i^c = 0.4\pm0.1$ mm, no structure forms even if cracks do.

\begin{figure}
    \begin{center}
        \includegraphics[scale=0.7]{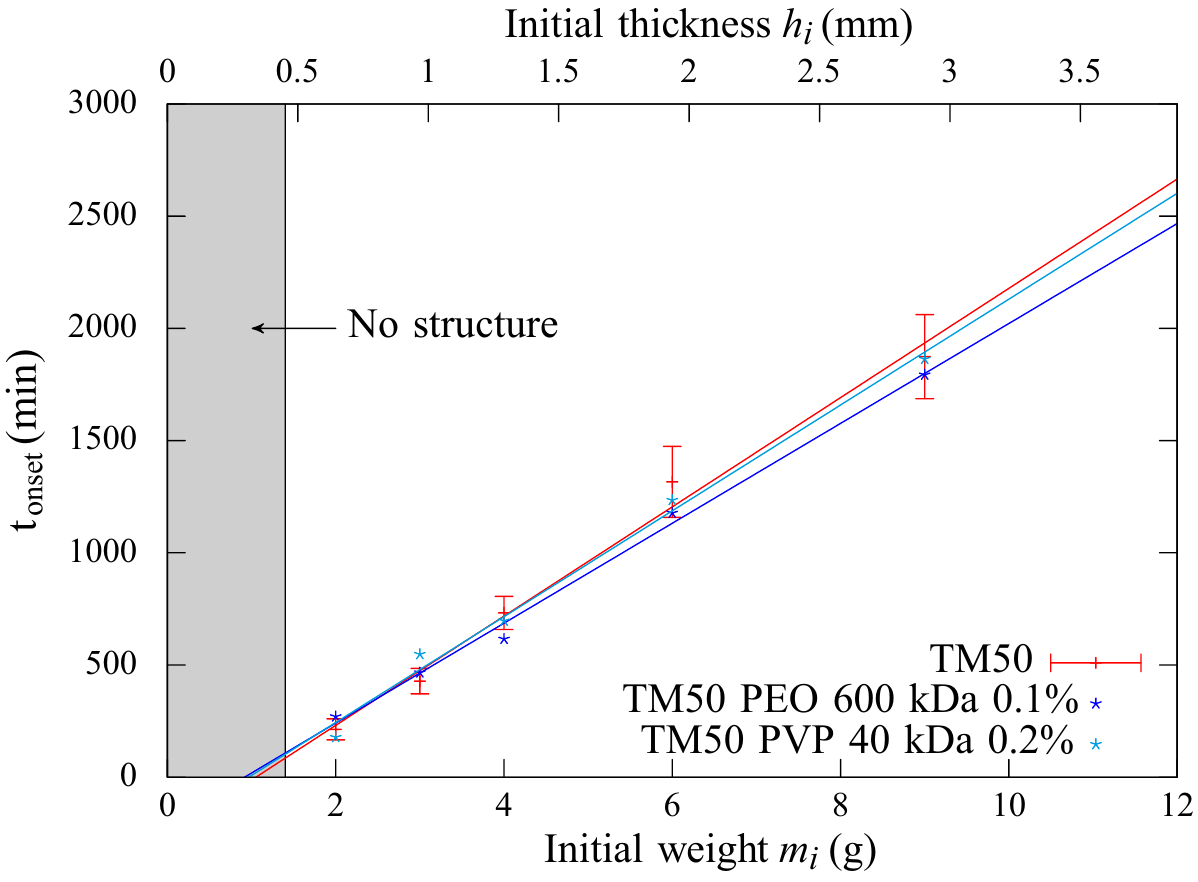}
    \end{center}
    \caption{Onset time $t_{onset}$ of structures as a function of the initial weight and corresponding initial film thickness, for three samples: TM, TM + $P_5$ (PEO, \SI{600}{\kilo\dalton}) at $C_p=0.1$\% and TM + $P_6$ (PVP, \SI{40}{\kilo\dalton}) at $C_p=0.2$\%.
        Lines are guides for the eye. Error bars reported for TM samples are similar for other samples.
    }\label{fig:time}
\end{figure}

In addition, above this critical initial weight, the onset time $t_{onset}$ needed to observe the structures growth, is measured for dispersions of different initial weights deposited in the container, ranging from $1$ to $9$ g.
Results are reported in figure \ref{fig:time} and show that the onset time increases linearly with the initial weight.

\begin{table*}
    \caption{Final state of sample surfaces for the different samples.}\label{tab:final_state}
    \centering
    \begin{tabular}{p{2.0cm}|p{1.7cm}||p{1.7cm}||p{1.7cm}||p{1.7cm}|p{1.7cm}:p{1.7cm}}
        \hline
        Sample & SM & HS & TM &  TM + ($P_0$, $P_1$, $P_2$, $P_3$) & \multicolumn{2}{c}{TM + ($P_4$, $P_5$, $P_6$, $P_7$)}   \\
        \cline{5-7}
        & & & & $\forall C_p$ & $C_p<C_p^c$ & $C_p>C_p^c$ \\
        \hline
        Observation & $\forall h$, no pattern & $\forall h$, no pattern  & patterns if $h>h_c$ & patterns if $h>h_c$ & patterns if $h>h_c$ & $\forall h$, no pattern  \\
        \hline
    \end{tabular}
\end{table*}

Similar experiments have been carried out with the other samples and results are summarized in table \ref{tab:final_state}.
For SM and HS samples, no structure appear at the film surface, independently of the investigated range of film thicknesses.

Consequently, we studied the addition of adsorbing polymers only on TM samples.
For a sample of initial weight $m_i=3$ g deposited in the container ($h_{i}=1$ mm), using PEO with molecular weights larger than $6$ kDa, the formation of the structures is suppressed above a critical concentration $C_p^c = 0.13 \%$ (within an uncertainty of $0.03\%$ for $P_3$ and $0.01\%$ for $P_4$, $P_5$ and $P_6$).
Thus, the weight amount of polymers necessary to suppress the structures is found to be independent of the molecular weight (for $M_w\geq6\textrm{kDa}$).
However, for shorter polymer chains ($M_w<6\textrm{kDa}$), and for concentrations up to $C_p=0.6$\%, the structures still form.

Moreover, addition of PVP ($40$ kDa) to TM particles results in suppression of the structures for $C_p^c(PVP) = 0.33\pm0.03$\%.
We notice that the molecular weight of a polymer unit for PEO $M_w^{PEO} = 44$ g/mol and for PVP $M_w^{PVP}=111$ g/mol:
\begin{equation}
    \frac{M_w^{PEO}}{M_w^{PVP}} \simeq \frac{C_p^c(PVP)}{ C_p^c(PEO)}
\end{equation}
As a result, structures are suppressed for a critical number of polymer units ($160\pm10$ units per colloidal particle, \textit{i.e.} $7.6\pm0.6$ \si{\micro\gram\per\meter\squared}).
We checked that drying kinetics are not affected by the addition of polymers by weight measurements \cite{Boulogne2013}.

\subsubsection{State of the material}
The drying of colloidal dispersion is a balance between two opposite fluxes: (a) the solvent flux tends to accumulate the particles to the free surface, whereas (b) the diffusion process smooths concentration gradients.
The competition between both processes usually states either the formation of a skin at the surface of a drying film or the consolidation in the bulk phase.
In that way, the relevant dimensionless number is a P\'eclet number defined as the ratio of a diffusion timescale $h^2/D$ and an advection timescale $h/V_e$: $\textrm{Pe}(h) =\frac{V_e h }{D}$ \cite{Routh1998},
where $h$ is the film thickness, $V_e$ is the evaporation rate and $D$ a diffusion coefficient.
The evaporation rate is found to be equal to $V_e \simeq 1\times 10^{-8}$m/s under the same drying conditions\cite{Boulogne2013}.
The diffusion coefficient is deduced from the Stokes-Einstein relation $D = k_B T/(6\pi \eta_s a)$, where $\eta_s =$ \SI{1}{\milli\pascal \second} is the solvent viscosity.
From thicknesses lying in the range $[0.4,2.9]$ mm, the P\'eclet number is between $0.2$ and $1.8$.
The value close to $1$ does not allow us to distinguish between skin formation or consolidation in the bulk.

However, to determine the state of the material, we gently collected a fraction of the material located at the surface of our samples (\textit{e.g.} TM sample, $m_i=9$ g, typically $1/4$ of the paste thickness); the sample was operated during the propagation of structures and before the apparition of cracks.
We measured a volume fraction of $\phi=38\pm2\%$ which is compatible with a elastic material ($\phi>0.35$) of non-aggregated particles ($\phi<0.61$) \cite{Boulogne2014a}; this material, deposited in a test tube filled thereafter with deionized water, can be redispersed after shaking.

\subsection{Visualisations}
\subsubsection{Optical microscopy}

\begin{figure}
    \begin{center}
        \includegraphics[scale=.5]{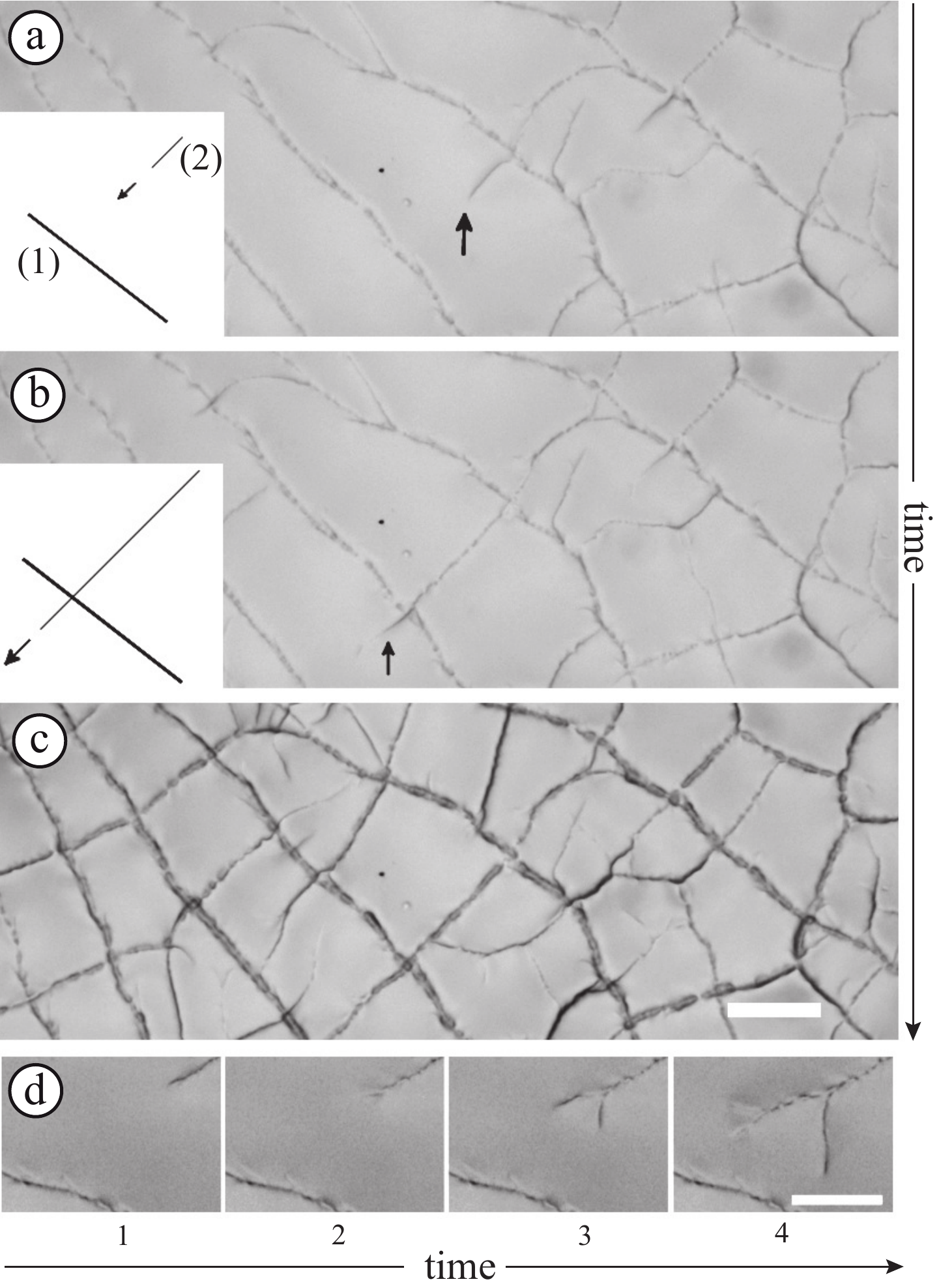}
    \end{center}
    \caption{Dynamics of the structure propagation in the center of a drying TM film.
        The time laps between pictures (a) and (b) is $21$ seconds.
        The situation indicated by arrows is detailed in the inserts.
        Picture (c) shows the final pattern.
        (d) The series of photographs show a morphogenetic sequence (time laps between each picture: $1$ s) leading to developed branching process.
        Scale bars represent \SI{200}{\micro\meter}.
    }\label{fig:opt_microscope}
\end{figure}

The propagation of structures was firstly observed by optical microscopy.
The sequence of images in figure \ref{fig:opt_microscope}(a, b, c) reveals the dynamics of formation of the structures in the center of the sample.
A propagating branch (2) approaches a prior one (1), then crosses.
Moreover, in figure \ref{fig:opt_microscope}(c), the final pattern in the sequence shows an increase of the structures contrast, and the formation of a new generation of branches, dividing domains.
The series of photographs in figure \ref{fig:opt_microscope}(d) show a typical detailed process of the splitting of one branch into two branches.


\begin{figure}
    \begin{center}
        \includegraphics[scale=1.0]{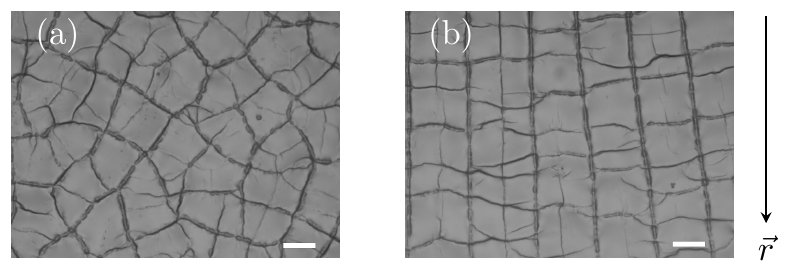}
    \end{center}
    \caption{
        Final pattern (a) in the center of the petri dish and (b) at $1$ cm from the edge located at the bottom of the picture (The vector $\vec{r}$ designates the radial direction.). Scale bars represent $100$ \si{\micro\meter}.
    }\label{fig:aniso}
\end{figure}
\begin{figure}[h!]
    \begin{center}
        \includegraphics[scale=.62]{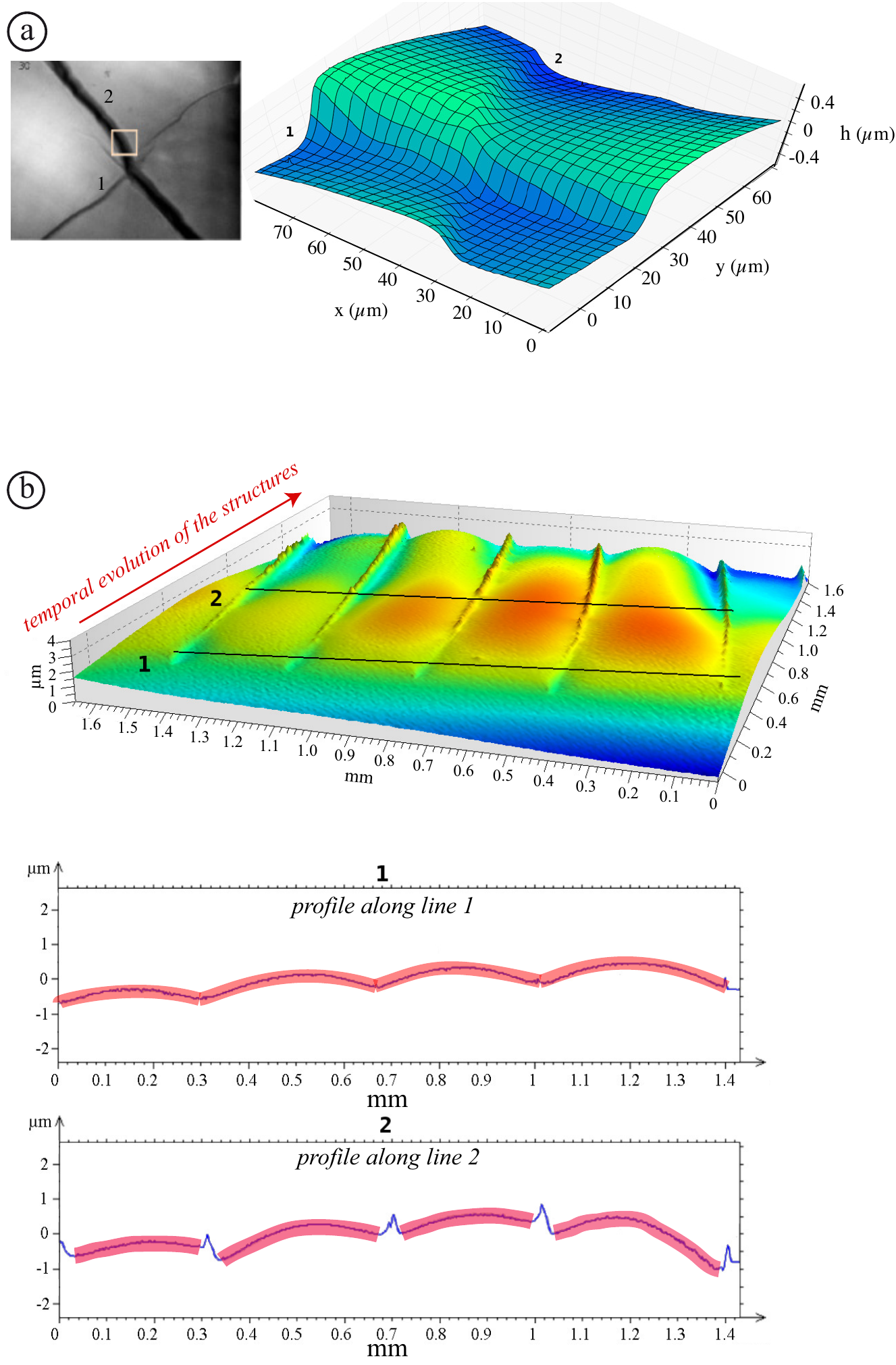}
    \end{center}
    \caption{(a) Surface profile (AFM) of a dried TM sample ($m_i = 6$ g). The 3D profile on the right corresponds to the region inside the yellow square in image on the left.
        (b) Surface profile obtained using an optical profiler: a drying region with propagating structures from the background to the foreground are shown in the profiles along the lines 1 and 2.
    }\label{fig:profilo}
\end{figure}

As shown in previous pictures, the structures exhibit an isotropic pattern in the center of the sample.
However, while structures propagate in a thickness gradient naturally imposed by the meniscus at the edge of the container, they become preferentially oriented radially and ortho-radially (figure \ref{fig:aniso}(b)).

A meniscus at the edge of the container extends over few times the capillary length scale $\kappa^{-1} = \sqrt{\frac{\gamma}{\rho g}} \simeq 2$ mm ($\gamma$ is the surface tension of water).
At the final stage, we obtain isotropic patterns (figure \ref{fig:aniso}(a)) in a region covering $40$\% of the total surface area which corresponds to the flat film area.


\subsubsection{AFM profilometry}

The surface topography of a TM sample is reported in figure \ref{fig:profilo}(a).
The typical structure height ranges between $1$ \si{\micro\meter} and $4$ \si{\micro\meter}.
In particular, we notice the asymmetric shape of the surface profile, as an evidence of two inclined planar surfaces connected by a transition region of about $10$ \si{\micro\meter}.
As a result, such measurements during the propagation of structures are shown in figure \ref{fig:profilo}(b).
The red region of the surface corresponds to a thicker layer; this can be related to the presence of the microscope objective limiting the evaporation flux \cite{Parneix2010}.
In front of the structures, the film is flat, and an arch-shaped profile is shown at the onset of the instability.
During the drying process, this shape evolves to an asymmetric shape.
The kinks present near the junction of arches (bottom graph) is due to the absence of interference fringes with the optical profiler at the singularities resulting in an artifact.
Thus, we only consider the  parts of the profile highlighted in red.

Moreover, in the case of experiments carried out with SM and HS samples for $m_i \in[2,12]$ g, no structure can be observed.
This statement is also confirmed using AFM, and optical profiler images.

Finally, measurements using AFM and optical profiler techniques reveal the absence of structures in TM with polymers above $C_p^c$ as in the case of observations using optical microscopy.
Note that polymer addition below $C_p^c$ leads to a lower structure height.
For instance, addition of PEO (\SI{600}{\kilo\dalton}, $C_p=0.1$\%) decreases the maximum height to \SI{1}{\micro\meter} ($m_i=6$ g).\\

\subsection{Rheology}
\begin{figure}
    \begin{center}
        \includegraphics[scale=0.75]{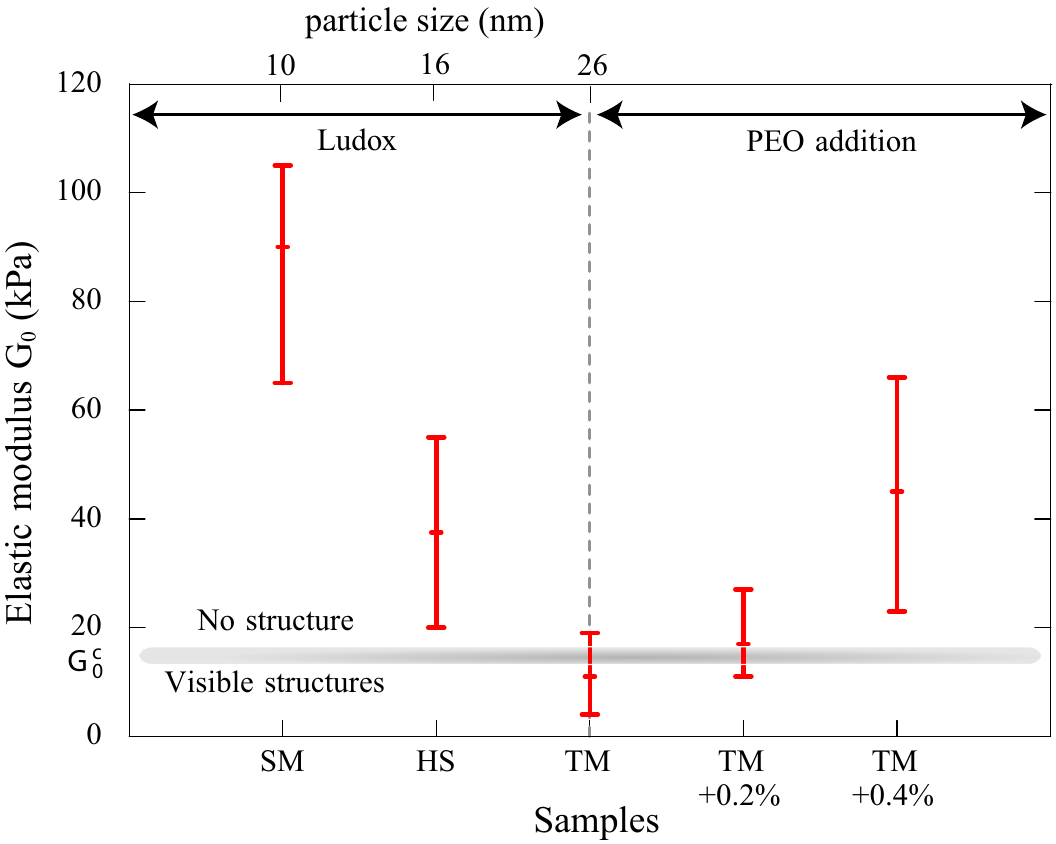}
    \end{center}
    \caption{Rheological measurements of the elastic modulus $G_0$ obtained from small oscillatory shear flow tests (plate-plate geometry) for different samples of pure particles (SM, HS and TM) and with addition of PEO ($600$ kDa) in TM. Error bars indicate extreme values.
    }\label{fig:rheo}
\end{figure}

In the following the rheological properties of pastes are investigated at the onset of structures formation.
In order to compare the elastic moduli of different silica pastes, measurements are performed at a well-defined consolidation time: the onset time for TM films and the same consolidation time for other systems.
Samples are prepared using the same protocol: the initial weights range from $7$ to $9$ g.
We collect surface samples corresponding to the 4/5th of the film thickness in the center of the container.
A part is used for a dried extract in order to deduce the volume fraction $\phi$.
The paste elastic modulus is measured using a rheometer (Anton Paar MCR501) in the plate-plate geometry (diameter: $24.93$ mm) with a solvent trap.
The gap is adjusted between $0.3$ and $0.7$ mm accordingly to the amount of paste used for the measurement.
Small oscillatory shear flow tests are performed with an amplitude of $0.2$\%, and a frequency sweep from $100$ Hz to $1$ Hz for $200$ s per point.
Under these conditions, the elastic modulus $G'(\omega)$ does not vary significantly with the frequency (relative variations lower than $10$\%) and the deformation amplitude up to few percent above which the elastic modulus decreases.
The average value is denoted by $G_0$ in the following.

In this way, five samples are considered: SM, HS and TM dispersions and binary mixture of TM and PEO (\SI{600}{\kilo\dalton}) at $C_p=0.2$\% and $C_p=0.4$\%.
For TM samples (without polymers), since structures appear at $\phi=38\pm2$\%, we select other samples (where no structures are developed) with similar volume fraction.
Results are shown in figure \ref{fig:rheo}.
Our results on the HS sample are consistent with values available in the literature\cite{DiGiuseppe2012}.
Note that, for all studied samples, the loss modulus ($G''(\omega)$) appears to be at least one order of magnitude lower than $G_0$.
For pure dispersions, the elastic modulus decreases of nearly one order of magnitude when the mean particle size varies from $10$ to $26$ nm.
On the contrary, the addition of a small amount of PEO adsorbing on TM particles results in the increase of the paste elastic modulus.

\section{Discussion}

Our observations concerning the dynamics of formation and the characteristics of the pattern show that the structures observed here can be scarcely related to winkles.
Indeed, wrinkles are periodic deformations of the interface\cite{Chen2012a} whereas our patterns are clearly localized.

Folds, described as a secondary instability arising from wrinkles\cite{Brau2013}, present strong similarities such as the localization of the deformation, the subdivision of domains and the crossing patterns\cite{Kim2011}.
However, the observation of wrinkles is expected but it has never been observed in our systems.
Moreover, wrinkles and folds result from the deformation of a skin in contact with a softer material whereas at the onset of the structures formation, the film seems to behave like an homogeneous paste phase.

Observations at the onset of patterning (top graph of figure \ref{fig:profilo}(b)) suggest some morphological similarities with creases patterns of swelling elastomers reported in the literature \cite{Cai2012} while the mechanism is different.
The observed arches evolve to a``saw-tooth roof shape'' as shown by AFM measurements (Fig. \ref{fig:profilo}(a)).
This transition is illustrated in the bottom graph of figure \ref{fig:profilo}(b) where the onset of the lost of symmetry can be seen.
Indeed, this transition may be attributed to the fact that colloidal pastes are able to change their microstructures.
A possible explanation of the symmetry breakage from the onset of the instability to the final state can be that any further strain energy is dissipated in a shear of the material (creep flow).

The second question addressed in this paper concerns the absence of structures for SM and HS particles, as well as the suppression by the addition of adsorbing polymers.

It results that mechanical properties of pastes depend on the colloidal dispersion or the concentration of additional polymers.
The strain in the paste phase at the onset time can be deduced its elastic modulus.
Indeed, assuming that the internal stress during the drying process comes from the flux of water through the porous matrix made by silica particles, the greatest stress occurs at the drying surface where the pore pressure is the larger\cite{Brinker1990}.
There, the stress $\sigma$ scales as follows:

\begin{equation}
    \sigma \sim \frac{h\eta_s}{3k} V_e
\end{equation}
where $k$ is the permeability of the porous material.
This expression justifies that below a critical thickness $h_i^c$, no structure are visible because the tensile stress is not large enough to trigger the mechanical instability.
Estimating the permeability from Carman-Kozeny relation at $\phi=38$ \% ($k=6\times 10^{-18}$ \si{\meter\squared}), we deduce a critical strain $\epsilon_c$ at the onset of the structures formation:
\begin{equation}
    \epsilon_c \sim \frac{h_i^c\eta_s}{3k G_0} V_e \simeq 0.2\label{eq:strain}
\end{equation}
where $G_0=11$ kPa for TM.
It is noteworthy that for this deformation, the elastic modulus of the material is still comparable to the one measured at low deformation amplitudes.
Moreover, the elastic modulus is measured just prior the propagation of the structures (that occurs over a timescale of 10 min) at a controlled volume fraction.
Thus, this elastic modulus must be considered as an order of magnitude that could be underestimated, which means that this critical strain could beoverestimated.

Measurements show that above a critical elastic modulus, structures are suppressed. This critical value is estimated by $G_0^c = 15\pm5 $ kPa from figure \ref{fig:rheo}.
Indeed, for pastes exhibiting larger elastic moduli, shrinkage induced by the drying is withstood by the paste network leading to a low strain.
Further shrinkage is frustrated by adhesion of the paste onto the substrate and results in tensile stress and so cracks formation.
However, for pastes exhibiting lower elastic moduli, the strain is larger leading to possible surface corrugations.
Equation \ref{eq:strain} also suggest that $G_0^c$ should depend on the film thickness if $\epsilon_c$ does not.
From our observations, we cannot conclude on a dependence of $G_0^c$ with the film thickness because near the threshold, the presence or absence of structures are difficult to be stated.
Indeed, the structures density decreases with the film thickness as shown in the inset of figure \ref{fig:time}.

In the case of pure colloidal dispersions (SM, HS, TM), the particle size affects the elastic modulus (Fig. \ref{fig:rheo}), consequently the permeability.
Indeed, fitting the elastic moduli as a function of the particle size with a power law, $G_0\propto a^p$, leads to values $p\in[-1,-3]$.
The inaccuracy comes from the large error bars and the little number of available samples.
Taking into account the particle size dependency in the permeability with Carman-Kozeny model, the deformation varies as $\epsilon \propto a^{-(p+2)}$.
A power coefficient $p$ lower than $-2$ would be consistent with our observations on the suppression effect.
We can also note that our description assumes that changing Ludox only results in varying particle size.
It is worth to note that other parameters may have effects, such as polydispersity (see reference \cite{Bergna1994} (p.~324) for measurements) which can modify the permeability \cite{Scherer1989b}.

Finally, we also observed that patterns are hierarchical (Fig. \ref{fig:timeline}) which introduces different lengthscales.
The final pattern settles after $250$ minutes in the given example.
This suggests that the refinement is related to the consolidation of the material that can be analyzed by the increase of the elastic modulus in time which increases the stress in the material.

\section{Conclusion}

In this paper, we report experimental observations on surface instabilities during the drying of silica colloidal dispersions.
These patterns grow if films are thicker than a critical thickness and that the onset time increases linearly with the initial film thickness and it is concomitant with a paste phase sitting on the substrate.
Measurements using microscopy techniques highlight the surface corrugations resulting first in arches that evolves to a saw tooth roof shape.
We also provide a method to prevent these patterns consisting in the addition of a small amount of polymers which increases the material elastic modulus.
From this technique, we estimated a critical elastic modulus for structures formation.

As a perspective, it would be interesting to refine the description of the evolution from arches to the saw-tooth roof shape which is attributed to a creep flow caused by the nature of the material.
Thus, a competition between the creasing instability and the creep flow might may exist in the selection of the final characteristic distance separating structures.
In particular, in a future work, the evolution of structures could be related to the characteristics (onset time, final size) of the different generations in order to establish a criterion for the distance between structures and to precise the influence of loss of symmetry in the development of the instability.
Further theoretical developments might focus on similarities and differences of surface instabilities in visco-elasto-plastic materials compared to visco-elastic gels.

\footnotesize{
\bibliography{structure} 
\bibliographystyle{rsc} 
}

\section{acknowledgments}

The authors thank Triangle de la Physique for the rheometer (Anton Paar, MCR 501) and A. Aubertin for designing the experimental setup.
We are grateful to B. Cabane, J.-P. Hulin, G. Gauthier and C. Quilliet for fruitful discussions, A. Chennevi\`ere, C. Poulard and F. Restagno for guidance and sharing their AFM and optical profiler.

\end{document}